\documentclass[a4paper,11pt,srt&compress]{article}
\usepackage{pos}
\usepackage{upgreek,xspace}

\def\textgamma{\ensuremath\upgamma}
\def\textsigma{\ensuremath\upsigma}

\def\textnu{\ensuremath\upnu}
\usepackage{multicol}
\usepackage{hyperref}

\title{Status of direct detection searches of heavy dark matter}

\author[*]{Marcin Ku\'zniak}

\affiliation{AstroCeNT, Nicolaus Copernicus Astronomical Center of the Polish Academy of Sciences,\\
  ul. Rektorska 4, Warsaw, Poland}

\emailAdd{mkuzniak@camk.edu.pl}
\note[*]{The speaker has been supported from the EU’s Horizon 2020 research and innovation programme
under grant agreement No 952480 (DarkWave), and from the International Research Agenda Pro-
gramme AstroCeNT (MAB/2018/7) funded by the Foundation for Polish Science from the European Regional Development Fund.}

\abstract{Dark matter accounts for 26\% of the mass-energy density of the Universe, however, its nature and origins remain the most important open questions in physics. The search for Weakly Interacting Massive Particles (WIMPs), one of the leading dark matter particle candidates, is now in a decisive phase, with experiments targeting both the high-mass and the low-mass (1--10~GeV/$c^2$) WIMP scenarios. The status of the leading experimental searches is discussed, together with their prospects and challenges they are facing. Searches of heavy non-WIMP dark matter candidates are also briefly summarized. 
}

\FullConference{ XVIII International Conference on Topics in Astroparticle and Underground Physics (TAUP2023)\\
 28.08~--~01.09.2023\\
University of Vienna\\}

\begin{document}
\renewcommand{\hookAfterAbstract}{%
\par\bigskip
\textsc{ArXiv ePrint}:
\href{https://arxiv.org/abs/2312.10828}{2312.10828}
}
\maketitle
\section{Introduction}
\subsection{Case for heavy dark matter}
The existence of dark matter (DM) is supported by a plethora of observational evidence. It constitutes some 26\% of the mass-energy balance of the Universe, is in the form of non-baryonic, non-dissipative and, most likely, cold DM, and yet its exact nature remains one of the main remaining puzzles in physics.
Among the various candidates, Weakly Interacting Massive Particle (WIMP) produced thermally via freeze-out, called standard or thermal WIMP, is one of the favoured DM candidates and received much focus over the last decades, alongside axions. Assuming electroweak strength interactions with the Standard Model sector and respecting the relic density constraints, implies order of GeV/$c^2$ to TeV/$c^2$ WIMP mass, potentially consistent with neutralino, the lightest stable particle expected from a large class of SUSY-based or other types of SM extensions (for a recent review, see e.g.~\cite{appec}).
Relaxing somewhat the assumptions on the interaction strength, the messenger particles involved and the DM production mechanism (thermal vs. not), shifts the lower mass constraint down to the eV/$c^2$ scale, with multiple possible DM scenarios beyond the scope of this report.
At the same time, super-heavy DM, with mass from multi-TeV/$c^2$ all the way up to the Planck scale (10$^{19}$~GeV/$c^2$), may result from more exotic scenarios, e.g. inflaton decay or gravitational mechanisms related to inflation (often related to Grand Unified Theories), millicharged gravitino DM, DM produced by primordial black hole radiation or extended thermal production in the dark sector~\cite{deapmimp}.

\subsection{Direct detection}
The principle of direct detection of heavy dark matter particles originated 40~years ago from the paper by Drukier and Stodolsky~\cite{drukier}, where the coherent enhancement of the neutrino elastic scattering cross-section on nuclei was first calculated, resulting in values much larger than previously anticipated, and within long-term experimental reach. In addition to establishing foundations for the CE\textnu NS measurement (only recently accomplished~\cite{cevns}), that work was soon generalized by Goodman and Witten~\cite{witten} to the case of WIMPs, which led to the first dark matter detector concepts~\cite{drukier2}, and the first experimental limits~\cite{avignone}.

Since then, more than 7 orders of magnitude of the WIMP-nucleon scattering cross-section parameter space have been experimentally probed. Ultimately, the sensitivity of non-directional direct searches is limited by irreducible backgrounds from solar and atmospheric neutrinos, the so-called neutrino fog~\cite{fog}.

\section{Direct detection techniques}
The experimental signatures of WIMP dark matter are single-site low-energy (keV) nuclear recoils, as well as the annual modulation of the dark matter signal, caused by changes in the velocity of Earth with respect to WIMP wind, and resulting changes in event rate above the energy threshold.

For the case of more exotic super-heavy dark matter candidates, multi-site nuclear recoil events or electronic recoil events are also a usable experimental signature, if sufficiently distinct from the expected backgrounds.

In essence, dark matter detectors seek for any measurable effects in the target medium caused by the recoil events depositing their energy. These typically include scintillation, ionization and heat. Sensitivity to more than one of these channels, on the one hand, comes at the cost of increased readout complexity, additional materials and associated radioactive impurities, on the other hand, provides stronger particle identification and background discrimination handles. As background events caused by instrumental effects or natural sources of radioactivity, (e.~g. traces of radioactive isotopes in the detector materials or its surroundings) are the main challenge, designing a competitive dark matter detector is a non-trivial balancing act.

\subsection{Crystal detectors}
A conceptually simple strategy of using an $\mathcal{O}$(100~kg) matrix of scintillating NaI(Tl) crystals viewed by photomultiplier tubes (PMT) was adopted by the DAMA/LIBRA experiment~\cite{dama}. This is also the case for ANAIS-112~\cite{anais}, SABRE~\cite{sabre} and COSINE-100~\cite{cosine}, which aim to cross-check the DAMA/LIBRA signal claim. In this configuration, nuclear recoil events (WIMP-like) are not distinguishable from electronic recoils events (background). COSINUS~\cite{cosinus} experiment is extending that approach by operating the sodium iodide crystals cryogenically as bolometers, with added phonon (temperature) readout, which will enable event-by-event discrimination between e-/\textgamma\ and nuclear recoil events.

The strategy of reading out multiple interaction channels has been routinely used by experiments using other types of crystals. For instance, EDELWEISS~\cite{edelweiss} uses cryogenic Ge, and CDMS/SuperCDMS~\cite{cdms} employs Ge and Si crystals, instrumented with ionization and phonon readout, for e-/\textgamma\ background discrimination. A fraction of crystals is operated at higher voltage with ionization sensed indirectly via Luke-Neganov phonon production, which enables reaching lower energy thresholds at the expense of background discrimination power.
The CRESST~\cite{cresst} collaboration operates cryogenic scintillating bolometers based on calcium tungstate or lithium aluminate crystals, which provide background discrimination capabilities based on the scintillation and phonon signal ratio.

Typically, large arrays, stacks or "towers" of individually calibrated crystals, often of mixed types, are operated in a single cryostat, providing additional capability to veto multi-scatter events. Their common feature is very low $\mathcal{O}$(10~eV) threshold, but also a difficulty in scaling-up to large target mass, due to material availability and cost.

\subsection{CCDs}
Silicon based charged coupled devices (CCDs) equipped with skipper readout have been developed and operated by DAMIC~\cite{damic} and SENSEI~\cite{sensei} collaborations. While essentially sensitive to ionization, dense arrays of CCD pixels provide a powerful way to identify particles and discriminate backgrounds based on the hit pattern and track topology. Due to low energy threshold and good background discrimination, this technology is competitive in the low-mass WIMP regime.
 
\subsection{Noble element detectors}
Over the past decade, liquid argon or xenon based detectors have dominated the sensitivity for WIMP masses $>$1~GeV/$c^2$. Some of the main reasons include the ease of purification and scale-up to multi-tonne target masses, high scintillation and ionization yields as well as powerful self-shielding and background discrimination capabilities, for Xe and Ar, respectively. Furthermore, Xe has competitive sensitivity for spin-dependent scattering, particularly on neutrons.

\paragraph{Time projection chambers}
TPCs, such as currently running XENONnT~\cite{xenonnt}, LZ~\cite{lz}, PandaX-4T~\cite{pandax4t} with LXe, or with LAr  DarkSide-50~\cite{ds50lm} (already decommissioned), or DarkSide-20k~\cite{ds20k} (currently under construction), register primary scintillation (S1) and secondary scintillation (S2) from drifted electrons, employing the e- drift for position reconstruction and fiducialization. The scintillation to ionization ratio is the main electronic recoil discrimination handle in LXe. Gaseous TPCs (although not only noble element based) are also the primary technology of choice for directional detection, providing position resolution better than the nuclear recoil track length.
\paragraph{Calorimeters}
Calorimeters such as XMASS~\cite{xmass} with LXe, or currently-running DEAP-3600~\cite{deapbgr} with LAr, rely exclusively on scintillation. A big advantage of this approach is simplicity and scalability. Uniquely in LAr, pulse-shape discrimination of S1 signals provides a powerful electronic recoil discrimination power, on the level of 10$^9$ at threshold~\cite{deappsd}. 

\subsection{Bubble chambers}
In bubble chambers process of nuclear recoil induced heat deposition, nucleation and bubble formation in overheated liquids is employed, with the thermodynamical conditions of the system optimized to suppress any events from electronic recoils. Fluorine-rich target liquids used by PICO~\cite{pico} give it leading sensitivity for spin-dependent (SD) interactions.
Scintillating Bubble Chamber (SBC)~\cite{sbc} has prototyped a detector combining the features of a LAr calorimeter and bubble chamber, with a potential to reach low energy thresholds.

\subsection{Paleodetectors}
Examining ancient minerals for traces of WIMP-nucleus interactions recorded over timescales as large as 1~Gyr may provide competitive exposure (and at least in principle, circumvent the limitations of the neutrino fog even without directional sensitivity). R\&D and small-scale demonstrations of this approach are ongoing~\cite{baum}.

\section{Status of experimental searches}
\subsection{DAMA/LIBRA detection claim}
Very significant efforts invested over the past decade into independent cross-check of the DAMA/LIBRA signal claim~\cite{dama} are now beginning to bring fruits. 

The claim has been already challenged by modulation searches in other targets. In LXe it was excluded by XENON100, LUX, and XMASS~\cite{xmass}. Recently, DarkSide-50~\cite{ds50mod} reported the first annual modulation search in LAr, with the lowest achieved energy threshold of 0.04~keV: compatible with no modulation, but neither confirming nor rejecting DAMA.

Searches using the same target material as DAMA are quickly catching up. COSINE-100~\cite{cosine} excluded most of the signal region with the model-dependent approach and reported a model-independent modulation search with 2.8~yr of data, consistent with both DAMA and no modulation; results from doubled exposure at a lower energy threshold are expected soon. ANAIS-112~\cite{anais} provided the first sensitive model-independent cross-check: incompatible with DAMA at $>$3.9\textsigma, consistent with no modulation; it is expecting to reach 5\textsigma\ sensitivity by the end of 2025.

\subsection{Low-mass WIMPs}
Recently improved S2-only DarkSide-50 analysis  resulted in the leading spin-independent (SI) sensitivity in the 1.2 to 10~GeV/$c^2$ range~\cite{ds50lm}, see Fig.~\ref{fig:sensitivity}~(top), with XENON1T, CDMSLite~\cite{cdms}, CRESST-III and DAMIC up to 2 orders of magnitude worse. Taking advantage of the Migdal effect to reduce the considered energy threshold extends the leading sensitivity of DarkSide-50 down to 0.04~GeV/$c^2$~\cite{ds50me}. Having said that, Migdal effect has not been experimentally confirmed yet~\cite{mexu} and its presence is, in principle, an open question investigated experimentally by multiple groups~\cite{me}. In the SD sector CRESST-III~\cite{cresst} holds the most stringent exclusion at 1~GeV/$c^2$ mass, while CDMSLite~\cite{cdmssd} and PICASSO~\cite{picasso} at 3~GeV/$c^2$ for scattering on neutron and proton, respectively. Beyond 3~GeV/$c^2$ XENON1T~\cite{xenon1t} and PICO-60~\cite{pico} take over.

\subsection{Standard thermal WIMPs}
\begin{figure}
    \centering
    \includegraphics[width=0.7\linewidth]{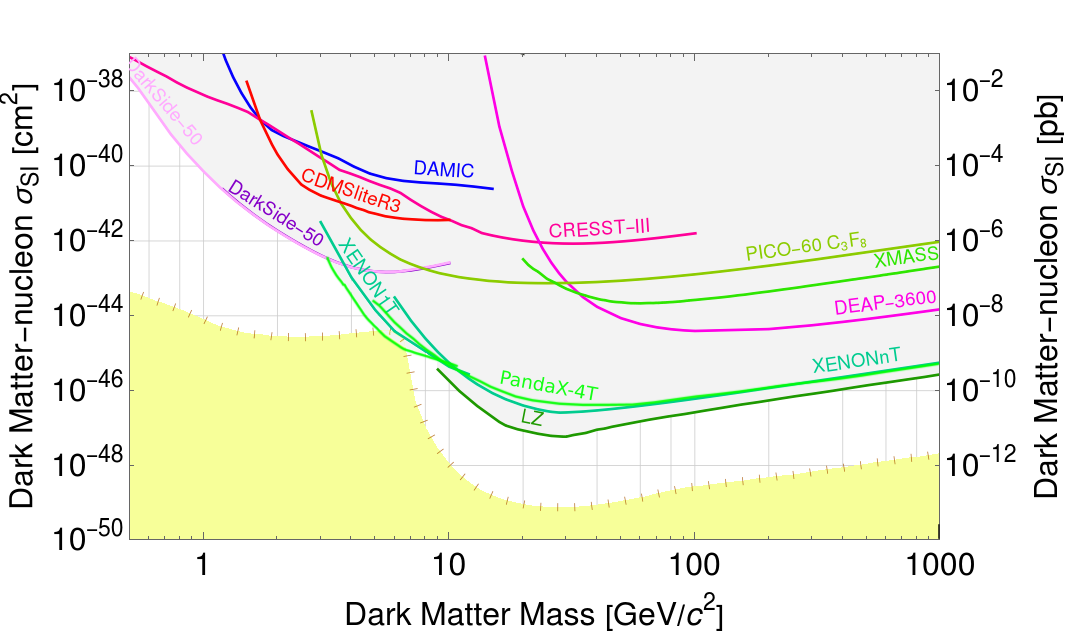}\\
    \includegraphics[width=0.5\linewidth]{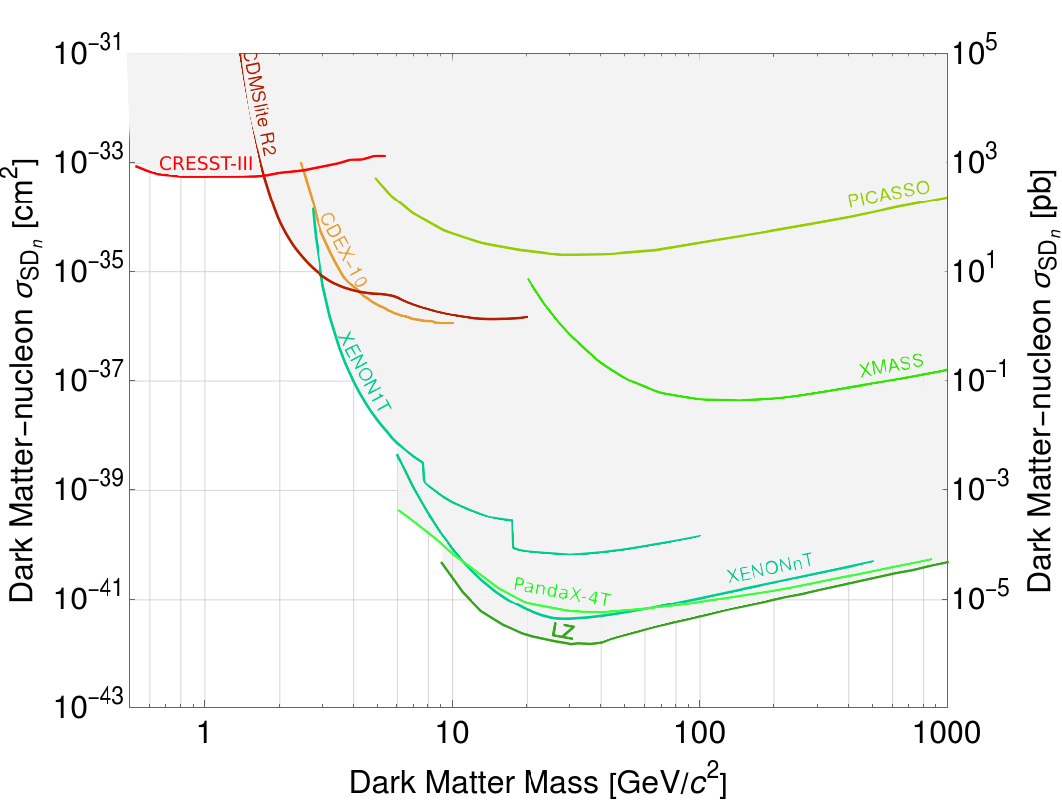}~~~~~~\includegraphics[width=0.51\linewidth]{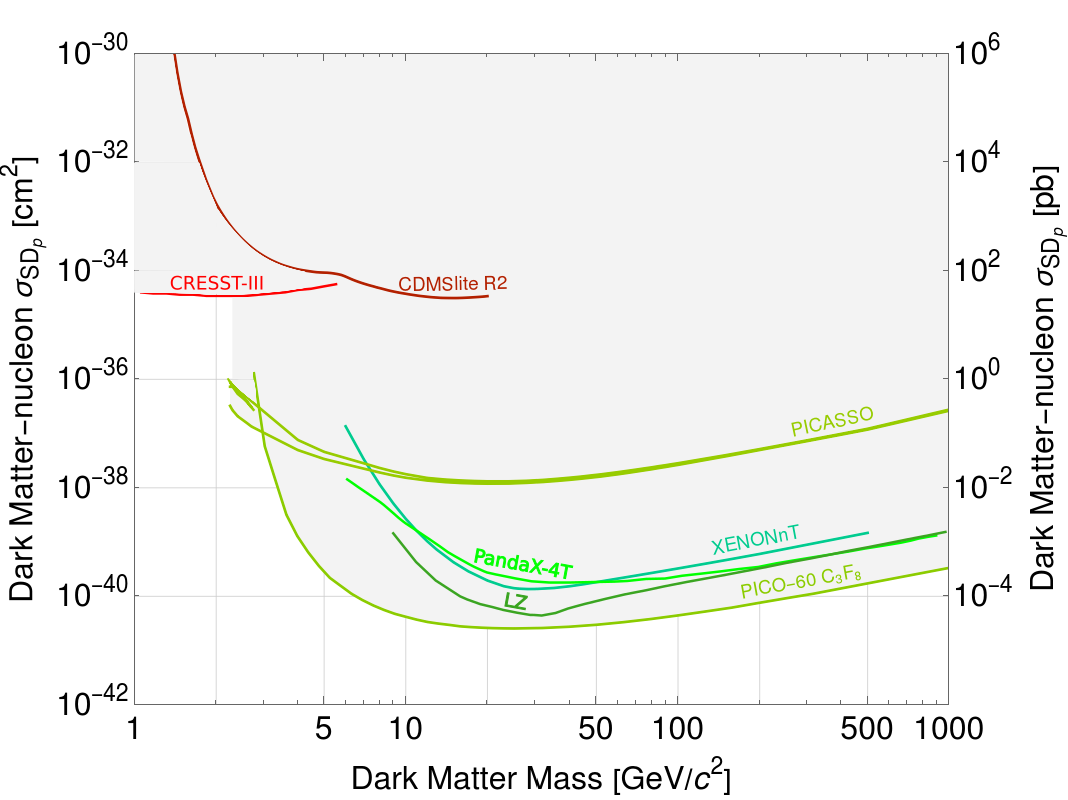}
    \caption{Status of direct detection searches: (top) spin-indepenent, (bottom left) spin-dependent WIMP-neutron scattering, (bottom right) spin-dependent WIMP-proton scattering. Generated with \href{https://supercdms.slac.stanford.edu/science-results/dark-matter-limit-plotter}{DM~Limit~Plotter}}
    \label{fig:sensitivity}
\end{figure}

In the high-mass regime, exposure is the critical factor and noble liquid detectors, primarily LXe detectors are unchallenged, see Fig.~\ref{fig:sensitivity}. Early results from the new generation of LXe experiments, XENONnT~\cite{xenonnt}, LZ~\cite{lz} and PandaX-4T~\cite{pandax4t}, have already pushed down the exclusion limits by more than a factor of 2 with respect to the previous generation of detectors. The only exception is the SD WIMP-proton scattering, where PICO-60~\cite{pico} has the leading edge, due to highly fluorinated target.
In case of LAr, the DEAP-3600 experiment is targeting reducing the sensitivity gap, following the ongoing upgrade addressing the limiting backgrounds~\cite{deapbgr}. 

\subsection{Planck-scale DM}
Due to a very distinct expected event signature, with high deposited energy and a large number of nuclear recoils all along the dark matter particle track in the detector, this type of searches is essentially background-free, with the sensitivity driven by the cross-sectional area of a detector. This gives the leading SI sensitivity at Planck-mass scale to DEAP-3600~\cite{deapmimp}, while at lower masses and cross-sections, or overall for the SD case, to XENON1T~\cite{xenonmimp}.

\section{Future prospects}
\paragraph{Low-mass WIMP region and low-energy excess}
Cryogenic bolometers and LAr TPCs are currently best positioned to push towards the neutrino fog for WIMP masses $<$10~GeV/$c^2$. For bolometers, a limiting factor currently is a low-energy event excess observed by multiple technologically different experiments (including DAMIC, SuperCDMS, CRESST)~\cite{excess}. The community has joined forces to address and resolve the issue in a collaborative manner. The mitigation strategy for such events and the exact origin of the excess remain open issues, with hints that energy release from long-lived metastable states is at play. A concept of DarkSide-LM, a tonne-scale LAr TPC optimized for S2-only operation and low mass sensitivity, has been proposed~\cite{dslm}, inspired by DarkSide-50. Also in this case, a low-energy excess was observed in DarkSide-50, and at least in part attributed to impurities, capturing electrons and releasing them with a delay. Success of this project depends on securing funding, and on the efficiency of the impurity removal and spurious electron event mitigation.

\paragraph{Directional searches}
Directional sensitivity allows to distinguish the expected WIMP signal from neutrino backgrounds, although so far it was only demonstrated in gaseous detectors with low target mass, and far from competitive exposure. Possibility to employ the columnarity effect in LAr to, at least partly, gain directional sensitivity in a scalable medium has been recently excluded~\cite{red}. However, the community has converged into the CYGNUS collaboration, with the goal of scale-up. Particularly, in the very-low WIMP mass regime, for WIMP-electron scattering, or for spin-dependent scenarios interesting sensitivities could be reached~\cite{cygnus}.
\paragraph{"Ultimate" large detectors}
Above 10~GeV/$c^2$, exposure of 200~tonne-year in LXe or 3000~tonne-year in LAr would be needed to reach the neutrino fog. Both communities have consolidated efforts, with LZ, XENON and DARWIN, merging into XLZD~\cite{xlzd} with targeted 60~tonne LXe mass, and independently PandaX-xT~\cite{pandaxxt} with $>$30 tonne mass.
The LAr community has merged into the Global Argon Dark Matter Collaboration, aiming to build a 300~tonne fiducial mass detector, ARGO~\cite{appec}, following the commissioning of DarkSide-20k. While technological challenges to resolve remain for both targets (and the community, guided by ECFA, has organized to address them~\cite{ecfa}), due to excellent pulse shape discrimination, in particular for events from elastic solar neutrino-electron scattering, LAr is in principle capable of achieving background-free exposure, which would result in a higher 5\textsigma\ discovery sensitivity~\cite{appec}.

\section{Summary}
Despite the lack of uncontested WIMP signal, the field has made a tremendous progress over the past few years, with a new set of exclusion limits from the current generation of detectors recently published. In addition to this, many important efforts are progressing, with the columnarity in LAr recently excluded, and the conclusive verification of the Migdal effect and a cross-check of the DAMA/LIBRA claim already on the horizon. While the roadmap towards the neutrino floor remains sound, significant challenges remain ahead, particularly in the context of understanding and mitigating the low energy backgrounds. Growing consolidation of efforts in the community gives many reasons for optimism and hope that 40 years into direct detection searches we now hawe most of the distance to the neutrino fog already behind us.



\end{document}